\documentclass[11pt]{article}

\usepackage{multirow}

\usepackage[nottoc]{tocbibind}

\usepackage{geometry}
\usepackage[affil-it, auth-sc]{authblk}
\usepackage[utf8]{inputenc}

\usepackage{graphicx}
\usepackage{amssymb}
\usepackage{amsthm}
\usepackage{amsmath}
\usepackage{color, soul}
\usepackage{verbatim}
\usepackage{lineno}
\usepackage{todonotes}
\presetkeys{todonotes}{inline, color=blue!30, size=\small}{}

\definecolor{darkred}{rgb}{0.9, 0.0, 0.0}
\definecolor{darkgreen}{rgb}{0.0, 0.5, 0.0}
\usepackage[pdftex,colorlinks,bookmarks,linkcolor=darkred, citecolor=darkgreen]{hyperref}

\usepackage[ampersand]{easylist}

\usepackage{bm}
\usepackage{feynmf}

\newcommand{\be}{\begin{equation}}
\newcommand{\ee}{\end{equation}}

\usepackage[sort&compress,numbers]{natbib}
\usepackage{doi}

\usepackage{slashed}

\newcommand{\order}{{\cal O}}
\newcommand{\nl}{\nonumber \\ }

\makeatletter
\newcommand{\vast}{\bBigg@{4}}
\newcommand{\Vast}{\bBigg@{5}}
\makeatother

\usepackage{eso-pic}

\allowdisplaybreaks

\begin{document}

%% \begin{fmffile}{feynmffile} 
%% \fmfcmd{%
%% vardef middir(expr p,ang) = dir(angle direction length(p)/2 of p + ang) enddef;
%% style_def arrow_left expr p = shrink(.7); cfill(arrow p shifted(4thick*middir(p,90))); endshrink enddef;
%% style_def arrow_left_more expr p = shrink(.7); cfill(arrow p shifted(6thick*middir(p,90))); endshrink enddef;
%% style_def arrow_right expr p = shrink(.7); cfill(arrow p shifted(4thick*middir(p,-90))); endshrink enddef;}

%% \fmfset{arrow_len}{3mm}
%% \fmfset{arrow_ang}{12}
%% \fmfset{wiggly_len}{3mm}
%% \fmfset{wiggly_slope}{75}
%% \fmfset{curly_len}{2mm}
%% \fmfset{zigzag_len}{1.5mm}
%% \fmfset{zigzag_width}{1mm}
%% \fmfset{dash_len}{1.5mm}

%% \fmfcmd{
%%     path quadrant, q[], otimes;
%%     quadrant = (0, 0) -- (0.5, 0) & quartercircle & (0, 0.5) -- (0, 0);
%%     for i=1 upto 4: q[i] = quadrant rotated (45 + 90*i); endfor
%%     otimes = q[1] & q[2] & q[3] & q[4] -- cycle;
%% }
%% \fmfwizard

\AddToShipoutPictureFG*{
    \AtPageUpperLeft{\put(-60,-60){\makebox[\paperwidth][r]{CALT-TH-2017-063}}}  
    \AtPageUpperLeft{\put(-60,-75){\makebox[\paperwidth][r]{FERMILAB-PUB-17-400-T}}}  
    }%

%\linenumbers

\title{\bf Power Corrections to the Universal Heavy WIMP-Nucleon Cross Section}

\date{January 25, 2018}

\author[1,2]{Chien-Yi Chen}
\affil[1]{Department of Physics and Astronomy, University of Victoria, Victoria, BC V8P 5C2 Canada \vspace{1.2mm}}
\affil[2]{Perimeter Institute for Theoretical Physics, Waterloo, ON, N2L 2Y5 Canada \vspace{1.2mm}}

\author[3,4]{Richard J. Hill}
\affil[3]{Department of Physics and Astronomy, University of Kentucky, Lexington, KY 40506, USA \vspace{1.2mm}}
\affil[4]{Fermilab, Batavia, IL 60510, USA \vspace{1.2mm}}

\author[5]{Mikhail P. Solon}
\affil[5]{Walter Burke Institute for Theoretical Physics, California Institute of Technology, Pasadena, CA 91125, USA \vspace{1.2mm}}

\author[6]{Alexander M. Wijangco}
\affil[6]{TRIUMF, Vancouver, BC, V6T 2A3 Canada \vspace{1.2mm}}

\maketitle

\begin{abstract}
WIMP-nucleon scattering is analyzed at order $1/M$ in Heavy WIMP
Effective Theory.
The $1/M$ power corrections, where $M\gg m_W$ is the WIMP mass,
distinguish between different underlying UV models with the same universal
limit
and their
impact on direct detection rates can be enhanced relative to naive expectations 
due to generic
amplitude-level cancellations at leading order.   The necessary one-
and two-loop matching calculations onto the
low-energy effective theory for WIMP
interactions with Standard Model quarks and gluons are performed
for the case of an electroweak SU(2) triplet WIMP,
considering both the cases of elementary fermions and composite scalars.  
The low-velocity WIMP-nucleon scattering cross section is evaluated and
compared with current experimental limits and projected future sensitivities. 
Our results provide the most robust prediction for electroweak triplet
Majorana fermion dark matter direct detection rates; for this case, a
cancellation between two sources of power corrections yields a small
total $1/M$ correction, and a total cross section close to the universal
limit for $M \gtrsim {\rm few} \times 100\,{\rm GeV}$.
For the SU(2) composite scalar, the $1/M$ corrections introduce
dependence on underlying strong dynamics. Using a leading chiral logarithm
evaluation, the total $1/M$ correction has a larger magnitude and uncertainty
than in the fermionic case, with a sign that further suppresses the total
cross section.
These examples provide definite targets for future direct detection experiments
and motivate large scale detectors capable of probing to the neutrino floor
in the TeV mass regime. 
\end{abstract}

\newpage

\section{Introduction} 

The WIMP paradigm remains a leading explanation for astrophysical dark
matter~\cite{Goodman:1984dc,Jungman:1995df,Bertone:2004pz,Feng:2010gw,Feng:2014uja,Arcadi:2017kky,Roszkowski:2017nbc}.  Null results at the LHC~\cite{ATLASSUSY,ATLASexotic,CMSSUSY,CMSexotic}
suggest that new physics is heavy
compared to masses of weak scale particles, $\sim 100\,{\rm GeV}$.  This
situation presents experimental challenges.  For example, at high-energy colliders
it is difficult to produce and detect on-shell heavy states that are coupled weakly
to the Standard Model.  Production cross sections are small and novel search strategies
are required to distinguish signal from background.  For the
$SU(2)_W\times U(1)_Y$ charged WIMPs
considered in this paper, with masses above the electroweak scale, 
detection prospects remain challenging at foreseeable
colliders~\cite{Aad:2013yna,CMS:2014gxa,Cirelli:2014dsa,Low:2014cba,Golling:2016gvc,ATLAS-CONF-2017-017}. 
Indirect searches for WIMP annihilation signals present a complementary set of
opportunities and 
experimental challenges, and introduce dependence on astrophysical
modeling~\cite{Hisano:2003ec,Cohen:2013ama,Fan:2013faa,Hryczuk:2014hpa,Bauer:2014ula,
  Bramante:2016rdh,Baryakhtar:2017dbj,Krall:2017xij}.
  Heavy particle techniques can be similarly applied to this
  case~\cite{Bauer:2014ula,Baumgart:2014vma,Ovanesyan:2014fwa,Baumgart:2017nsr}.

The heavy WIMP regime is also challenging for direct detection prospects.
First, since the abundance of astrophysical dark matter particles for a given
local energy density scales inversely as the particle mass,
WIMPs are less abundant and detection rates for a given cross section 
are smaller.
Second, as the mass spectrum of new physics states
becomes stretched above the weak scale, the absence of accessible
intermediate states forbids the simplest higgs-mediated interactions
of WIMPs with nucleons, causing cross sections to be smaller.

However, although the interaction rates between WIMPs and nucleons may
become smaller, they also become more certain.  Heavy WIMP symmetry
emerges in the limit that the WIMP mass, $M$, is large compared to
the electroweak scale, i.e., $M \gg m_W$.  Scattering cross sections 
become universal for given WIMP gauge quantum numbers,
independent of the detailed UV physics~\cite{Hill:2011be,Hill:2013hoa}. 
For example, the cross section in this limit is independent of whether
the particle is scalar or fermion, composite or fundamental.
This universality provides robust sensitivity targets for
ambitious next generation direct detection experiments, and will
be key to interpreting any confirmed signal. 

In previous work, two of the authors (RJH and MPS) analyzed
the universal heavy WIMP limit for WIMP-nucleon
scattering~\cite{Hill:2011be,Hill:2013hoa,Hill:2014yka,Hill:2014yxa}. 
In this limit a generic amplitude-level cancellation~\cite{Hisano:2011cs,Hill:2011be,Hill:2013hoa}
was shown to suppress the low-velocity WIMP-nucleon cross section to the level of
$\sim 10^{-47}\,{\rm cm}^2$ for wino-like WIMPs (i.e., self-conjugate electroweak triplets),
and higgsino-like cross sections to an even smaller value.
It is natural to ask whether in the presence of such
cancellations, formally subleading effects can become numerically
relevant beyond naive dimensional estimates.
For example, focusing on the electroweak triplet case,
the cancellation results in a total amplitude whose magnitude is
$\sim 20\%$ the size of the component subamplitudes~\cite{Hill:2014yxa},
and a WIMP-nucleon cross section that is therefore suppressed by more than
an order of magnitude. 
For TeV scale WIMPs, corrections of order $m_W/M$ could potentially
enter at a similar numerical level. Here we analyze such $1/M$ power
corrections, and quantify the corresponding violations of heavy WIMP universality. 

The remainder of the paper is structured as follows.  
Section~\ref{sec:formalism} extends Heavy WIMP Effective Theory (HWET)
to incorporate $1/M$ power corrections, and
Sec.~\ref{sec:lowEFT} matches to the low energy effective
theory after integrating out weak-scale particles. 
Section~\ref{sec:numerics} computes the low-velocity scattering cross section
of WIMPs on nucleons.
Section~\ref{sec:summ} provides a summary and outlook. 

\section{Heavy WIMP Effective Theory at order {\boldmath $1/M$} \label{sec:formalism}}

Heavy particle effective theory can be used to analyze Standard Model (SM) extensions consisting
of electroweak multiplets whose mass $M$ is large compared to SM particle masses, $M \gg m_W$.
Additional heavy multiplets, of mass $M^\prime$,
may be integrated out for generic mass splitting
$M^\prime - M = \order(M)$.
The special case $M^\prime - M = \order(m_W)$ requires that the additional multiplet
appear explicitly in the HWET~\cite{Hill:2013hoa}.%
\footnote{
For a related application of heavy particle effective theory to the case of
an electroweak singlet bino that is nearly degenerate with a stop, see Ref.~\cite{Berlin:2015njh}.
  }
Here we focus on a single multiplet of self-conjugate heavy particle
fields with arbitrary spin, transforming under irreducible representations of
electroweak $SU(2)_W \times U(1)_Y$. Where a specific representation is required, 
we illustrate with an electroweak triplet.

Working through order $1/M$, the gauge- and Lorentz-invariant lagrangian
in the one-heavy-particle sector (i.e., bilinear in $h_v$)
is~\cite{Hill:2011be}
\begin{align}\label{eq:HWlagrangian}
  {\cal L} &= \overline{h}_v \bigg\{ iv\cdot D - \delta m - {D_\perp^2\over 2M}
  + c_H {H^\dagger H\over M}
  + c_{W1}  {\sigma^{\mu\nu} W_{\mu\nu} \over M}
  + c_{W2}  {\epsilon^{\mu\nu\rho\sigma} \sigma_{\mu\nu} W_{\rho\sigma} \over M}
  +  \dots \bigg\} h_v \,,  
\end{align}
where the timelike unit vector $v^\mu$ defines the heavy WIMP velocity,  
$D_\mu = \partial_\mu - ig_1 Y B_\mu - ig_2 W^a_\mu t^a$ is
the covariant derivative,
$W_{\mu\nu} = i[D_\mu,\,D_\nu]/g_2 =  W^a_{\mu\nu} t^a$ is the
field strength, and $D_\perp^\mu = D^\mu - v^\mu v\cdot  D$. 
The heavy particle field $h_v$ satisfies projection relations as discussed
in detail in Ref.~\cite{Heinonen:2012km}; for example, a fermionic heavy particle field
obeys $\slashed{v} h_v = h_v$.
The self-conjugate
condition is enforced in the effective theory by requiring invariance of the lagrangian under
\begin{align}\label{eq:conjugate}
v^\mu \to - v^\mu \,, \quad h_v \to h_v^c \,,
\end{align}
where $h_v^c$ denotes charge conjugation.
For an irreducible representation of a self-conjugate
field, we necessarily have zero hypercharge and integer isospin.
The interactions labeled by $c_{W1}$ and $c_{W2}$ are present
for the fermionic case. They contribute only to spin-dependent
interactions at low velocity and will be ignored in the following.

The coefficient, $-1/2$, of the kinetic term $D_\perp^2/ M$ in Eq.~(\ref{eq:HWlagrangian}) 
is fixed by relativistic invariance~\cite{Luke:1992cs,Heinonen:2012km}.
The residual mass, $\delta m$ in Eq.~(\ref{eq:HWlagrangian}), may be chosen
for convenience.  In a theory without electroweak symmetry breaking, taking
$\delta m=0$ would enforce that $M$ is the physical particle (pole) mass.
For matching calculations at the electroweak scale, it is convenient to choose
$\delta m = c_H  \langle |H|^2 \rangle /  M$
to cancel the mass contribution from electroweak
symmetry breaking.

\begin{figure}[t]
\begin{center}
\includegraphics{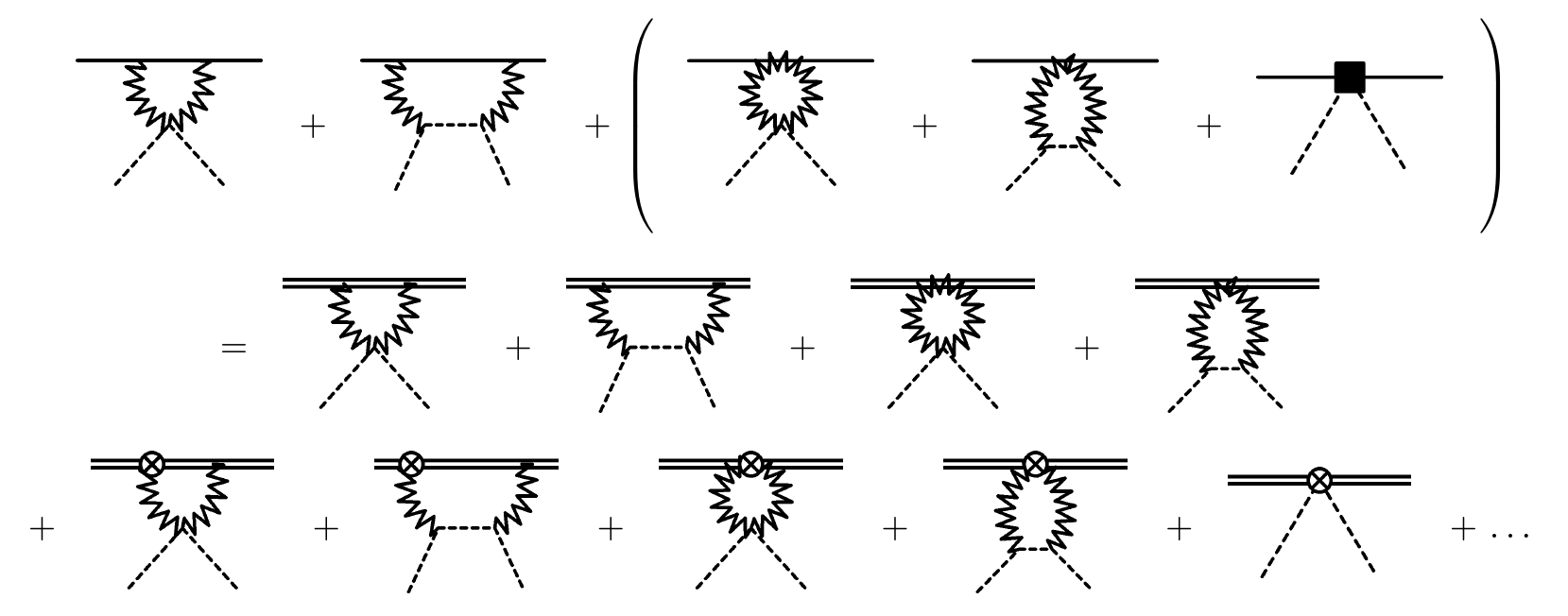}
  \caption{\label{fig:chmatch}
  Matching condition for the coefficient $c_H$ for UV theory consisting of
  the Standard Model plus $SU(2)_W$-triplet Majorana fermion.
  Solid lines denote Majorana fermion, dashed lines denote SM Higgs doublet,
  zigzag lines denote $SU(2)_W$ gauge fields.  Matching is performed in the
  electroweak symmetric theory.  
  Double lines on the RHS denote heavy WIMPs and the encircled cross denotes
  insertion of a $1/M$ effective theory vertex.
  For UV theory consisting of a composite real scalar transforming as
  a triplet under $SU(2)_W$, the additional bracketed terms appear on the LHS,
  including the counterterm contribution denoted by the solid square.
  }
 \end{center}
\end{figure}

The parameter $c_H$ encodes ultraviolet
physics above the scale $M$,
and can be determined by a matching computation between a specified UV theory and HWET,
described by Eq.~(\ref{eq:HWlagrangian}). 
As an example, let us consider the case where the UV theory is given by the SM
and an electroweak triplet of Majorana fermions.
Matching onto HWET is illustrated in Fig.~\ref{fig:chmatch}.  The matching can be
performed in the electroweak symmetric theory.  After expanding in the Higgs mass parameter,
the EFT diagrams are scaleless but dimensionful and thus vanish in dimensional regularization.
Evaluation of the full theory diagrams yields the matching condition,
\begin{align}\label{eq:cH_T}
  c_H(\rm Majorana\,\,fermion) = -3 \alpha_2^2 \,.
\end{align}
As a simple renormalizable extension of this case, consider an additional electroweak
multiplet transforming with higgsino quantum numbers (SU(2)$_W$ doublet, hypercharge $Y=1/2$) with mass $M_D$.
For generic doublet-triplet mass splitting, $M_D-M_T = \order(M_T)$,
the matching coefficient becomes
\begin{align}\label{eq:cH_TD}
  c_H({\rm doublet-triplet}) = - 3\alpha_2^2 + 4\pi \alpha_2 \kappa^2 {M_T \over M_D-M_T} \,,
\end{align}
where $\kappa$ is the  renormalizable trilinear coupling between the triplet and doublet fermions
and the SM Higgs field~\cite{Hill:2013hoa,Hill:2014yka}. 
As expected, when $M_D/M_T \to \infty$, the result (\ref{eq:cH_TD}) reduces to the pure
triplet result (\ref{eq:cH_T}).

As an example involving scalar versus fermionic WIMP, consider
the pseudo-Goldstone bosons that emerge from a QCD-like SM extension with
vector-like SU(2)$_{W}$ couplings to underlying
fermions~\cite{Kilic:2009mi,Bai:2010qg}.
Recall that the lightest such states form an electroweak triplet,
regardless of the fermionic SU(2)$_{W}$ representation, and 
these ``weakly interacting stable pions'' are stabilized by a discrete
symmetry (the unbroken analog of Standard Model G parity)~\cite{Bai:2010qg}. 
The matching is again illustrated in Fig.~\ref{fig:chmatch}, where now the
full theory diagrams involve relativistic scalars, and also a counterterm four point
function between the WIMP and SM Higgs field.
The one-loop diagrams are UV divergent
as a function of the cutoff $\Lambda_h$ representing the new strong interaction scale. 
The divergence is cancelled by the
counterterm contribution.  For the composite theory under consideration, 
the divergence corresponds to a logarithmically enhanced term in the matching.
Taking this ``chiral'' logarithm as an estimate, we have 
\begin{align}\label{eq:cH_S}
  c_H({\rm composite\,\,scalar}) = \alpha_2^2  \log{\Lambda_h^2\over M^2 }  + \dots 
    \approx \alpha_2^2  \log{1\over \alpha_2}  + \dots \,,
\end{align}
where the ellipsis denotes $\order(1)$ terms that are not logarithmically enhanced.
The last equality corresponds to a chiral symmetry breaking mass $M$ induced
by SU(2)$_{W}$ radiative corrections: 
$M^2/\Lambda_h^2 \sim \alpha_2$~\cite{Bai:2010qg}.  The precise matching condition could
in principle be computed using strong interaction methods in the chosen UV theory. 

The cases (\ref{eq:cH_T}), (\ref{eq:cH_TD}), and (\ref{eq:cH_S}) establish the range
of $c_H$ encountered in a variety of weakly coupled UV models, involving fermions and
scalars, composite and elementary particles, and both pure-state and multi-component models.
Before investigating the impact of these differences on direct detection cross sections,
let us perform the remaining step of matching HWET onto effective QCD operators.

\section{Effective Theory Below the Weak Scale \label{sec:lowEFT}}

The scale separation $m_W \gg \Lambda_{\rm QCD}$, is exploited by matching
onto a heavy particle effective theory for the relevant electrically neutral component of the
WIMP, interacting with five flavor QCD:
\begin{align}\label{eq:weaklagrangian}
{\cal L} = \overline{h}^{(0)}_v h^{(0)}_v
  \bigg\{ \sum_{q=u,d,s,c,b} \bigg[ c_q^{(0)} O_q^{(0)} + c_q^{(2)} v_\mu v_\nu O_q^{(2)\mu\nu} \bigg]
    + c_g^{(0)} O_g^{(0)} + c_g^{(2)} v_\mu v_\nu O_g^{(2)\mu\nu}
    \bigg\} + \dots \,. 
\end{align}
This matching step is common to different UV realizations of the electroweak triplet WIMP. 
In Eq.~(\ref{eq:weaklagrangian}),
$h^{(0)}_v$ is the neutral WIMP, and 
the spin-0 and spin-2 QCD operators for quarks and gluons are given by
\begin{align}
  O^{(0)}_{q} &= m_{q} \bar{q}q \,, &
  O^{(2) \mu \nu }_{q} &= \frac{1}{2}\bar{q}\left(\gamma^{\{\mu}iD_{-}^{\nu\}}-\frac{g^{\mu \nu}}{d}i\slashed{D}_{-}\right)q \,,
  \nl
  O^{(0)}_{g} &= (G^{A}_{\mu\nu})^{2} \,, &
O^{(2) \mu \nu}_{g} &= -G^{A\mu\lambda}G^{A\nu}_{\lambda}+\frac{1}{d}g^{\mu\nu}(G^{A}_{\alpha\beta})^{2} \,,
\label{eq:op}
\end{align}
where $d=4-2\epsilon$  is the spacetime dimension,
$D_{-} \equiv \overrightarrow{D} - \overleftarrow{D}$,
and curly braces denote symmetrization, 
$A^{\{\mu}B^{\nu\}} \equiv (A^{\mu}B^{\nu}+A^{\nu}B^{\mu})/2$.
The ellipsis in Eq.~(\ref{eq:weaklagrangian})
denotes higher dimension operators suppressed by $\Lambda_{\rm QCD}/m_W$,
and spin-dependent operators. 

By restricting to dimension seven operators in Eq.~(\ref{eq:weaklagrangian}),
we are neglecting contributions suppressed by additional powers of
$\Lambda^2_{\rm low-energy}/m^2_W$, where $\Lambda_{\rm low-energy}$ denotes
any scale below $m_W$ (e.g., $m_{b}$, or $\Lambda_{\rm QCD}$).
However, we will account for corrections of order $m_W/M$ in the coefficient
functions appearing in Eq.~(\ref{eq:weaklagrangian})
in our analysis of HWET power corrections. 
This power counting is appropriate for dark matter masses in
the few hundred GeV to TeV range, a focus for current and next generation
direct detection experiments. 

\begin{figure}[t]
\begin{center}
\includegraphics{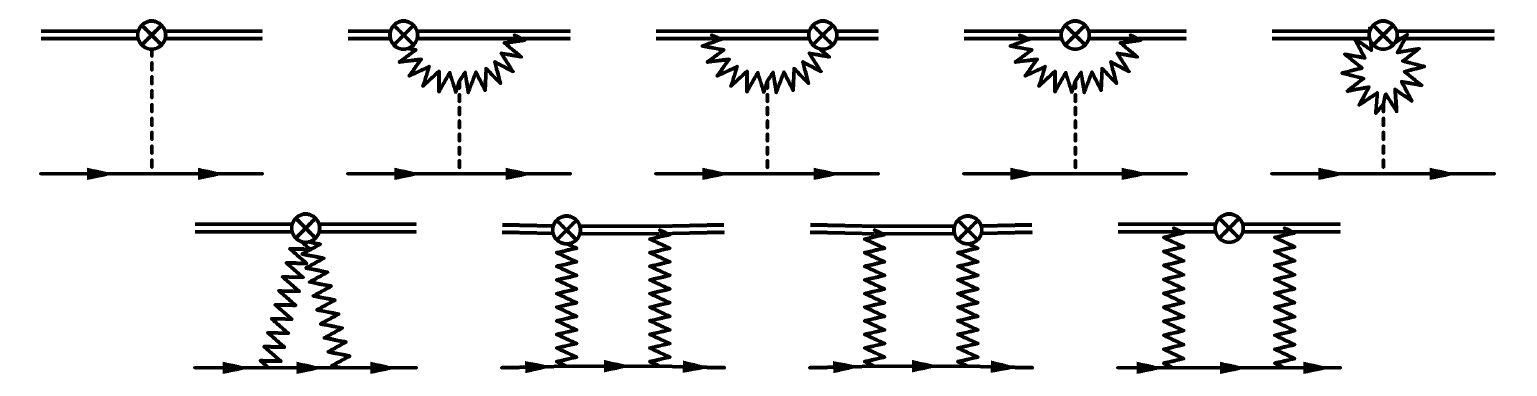}
  \caption{\label{fig:1Mquarkmatching}
  Diagrams contributing to $1/M$ quark matching, with the same notation as in
  Fig.~\ref{fig:chmatch}.  Diagrams with crossed $W$ lines are not displayed. 
}
\end{center}
\end{figure}

\begin{figure}[t]
\begin{center}
\includegraphics{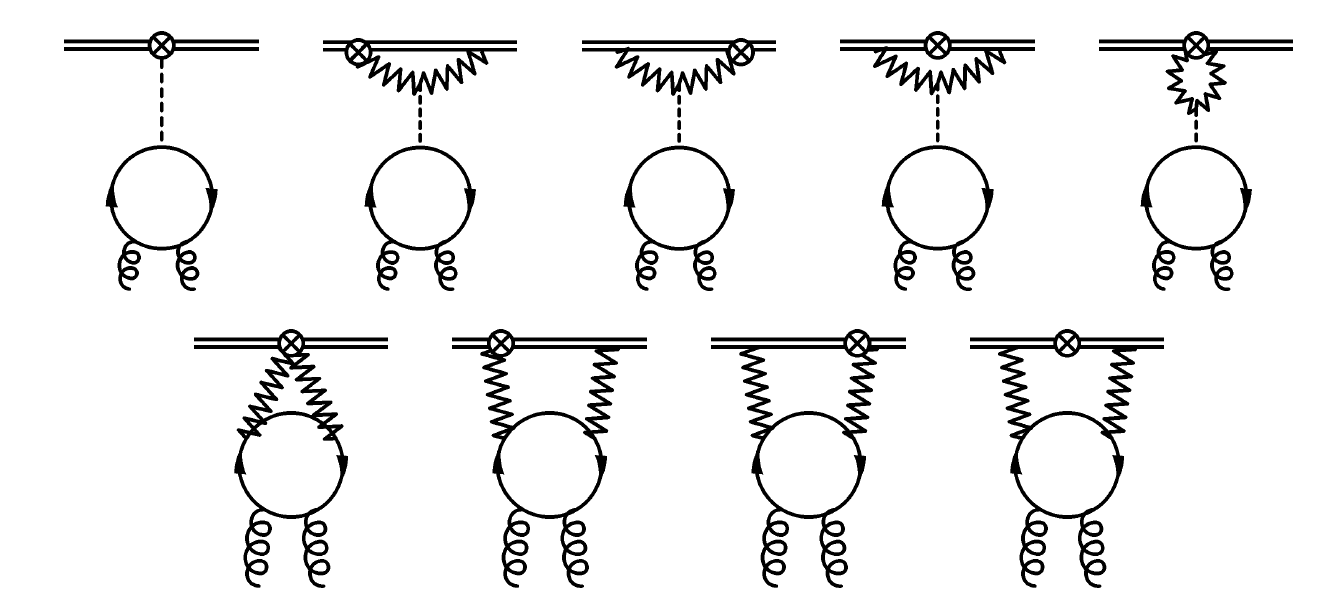}
  \caption{\label{fig:gluonmatch} Diagrams contributing to $1/M$ gluon matching,
  with the same notation as in Fig.~\ref{fig:chmatch}. Curly lines denote gluons. Diagrams with both gluons attached to the upper quark line or with one gluon attached to each of the upper and lower quark lines are not shown.
}
\end{center}
\end{figure}

We now proceed to match the theory (\ref{eq:HWlagrangian}) to the theory
(\ref{eq:weaklagrangian}). 
By integrating out weak scale particles (the Higgs boson, electroweak
gauge bosons, and the top quark), we obtain a solution for the twelve
effective theory coefficients
($c_q^{(0)}$ and $c_q^{(2)}$ with $q=u,d,s,c,b$, as well as $c_g^{(0)}$ and $c_g^{(2)}$) that specify the interactions of DM
with five flavor QCD.
We neglect subleading corrections involving light quark masses, and
use CKM unitarity to simplify sums over quark flavors. Approximating
$|V_{tb}| \approx 1$, these simplifications imply that
$c_u^{(S)} = c_d^{(S)} = c_{s}^{(S)} = c_c^{(S)}$ for $S=0,2$, 
leaving six independent coefficients. In the
following, we denote generic up- and down-type quarks in five-flavor QCD
by $U$ and $D$,
respectively, and an arbitrary quark flavor by $q$. 

Feynman diagrams contributing to the matching at $\order(1/M)$ for the
quark and gluon coefficients are shown in Figs.~\ref{fig:1Mquarkmatching}
and~\ref{fig:gluonmatch}, respectively. 
Diagrams for gluon operators contain an additional loop compared
to diagrams for quark operators.  However, owing to the large gluon matrix
elements of the nucleons, these operators are numerically of similar
size, or dominant.
We compute each of the operator coefficients in Eq.~(\ref{eq:weaklagrangian})
to leading order in electroweak couplings, and hence we neglect one-loop diagrams involving $c_H$ for quark matching and two-loop diagrams involving $c_H$ for gluon matching. The impact of higher order
contributions is estimated in the numerical analysis by varying
the factorization scale.
The techniques for electroweak scale matching detailed in Ref.~\cite{Hill:2014yka}
can be applied to the present calculation.
We describe some pertinent details here. 
Compared to the leading power
analysis considered in Ref.~\cite{Hill:2014yka}, computation of the $1/M$ corrections
requires an extended master integral basis, and different components of the
electroweak polarization tensor for the background field gluon matching.  

In performing the gluon matching, it is convenient to distinguish between
amplitudes with one or two bosons exchanged in the $t$-channel. One-boson
exchange amplitudes are shown in the top row of Fig.~\ref{fig:gluonmatch},
while two-boson exchange amplitudes are shown in the bottom row.
The one-boson exchange amplitudes factorize into the one-boson exchange
amplitudes for quark matching (top row of Fig.~\ref{fig:1Mquarkmatching})
times the quark loop, and contribute only
to the scalar coefficient. For the two-boson exchange amplitudes, we employ electroweak
polarization tensors, $\Pi^{\mu \nu}$, induced by a loop of quarks in
a background field of external gluons~\cite{Novikov:1983gd,Hisano:2010ct,Hill:2014yka}.
The temporal components, $v_\mu v_\nu
\Pi^{\mu \nu}$, are sufficient for the leading power analysis, while 
for the $1/M$ corrections we require also the spatial components;
these may be extracted from Ref.~\cite{Hill:2014yka}.
The renormalization of Wilson coefficients for the quark and gluon operators is
discussed in Ref.~\cite{Hill:2014yxa}.

From the sum of one and two loop diagrams in Figs.~\ref{fig:1Mquarkmatching}
and~\ref{fig:gluonmatch},
we obtain the final results for coefficients renormalized in the $\overline{\rm MS}$ scheme: 
\begin{align}\label{eq:results}
  \hat{c}_{U}^{(0)}(\mu)&= -\frac{1}{x_{h}^{2}} - \frac{m_W}{\pi M} \frac{c_{H}}{\alpha_{2}^{2} x_h^2}  \,,
  \nl
  \hat{c}_{D}^{(0)}(\mu)&= -\frac{1}{x_{h}^{2}} -\delta_{Db} \frac{x_{t}}{4(x_{t}+1)^{3}} 
 - \frac{m_W}{\pi M} \frac{c_{H}}{\alpha_{2}^{2} x_h^2} \,,
  \nl
  \hat{c}_{g}^{(0)}(\mu)&= \frac{\alpha_s}{4\pi}\bigg[ 
    \frac{1}{3x_{h}^{2}}
    + \frac{N_\ell}{6}
    +\frac{1}{6(x_{t}+1)^{2}}
  +\frac{m_W}{\pi M} \frac{c_{H}}{3\alpha_{2}^{2} x_h^2}
  \bigg] \,,
  \nl
  \hat{c}_{U}^{(2)}(\mu)&= \frac{2}{3}-\frac{m_{W}}{\pi M} \,,
  \nl
  \hat{c}_{D}^{(2)}(\mu)&= \frac{2}{3} + \delta_{Db}
  \left[\frac{3x_{t}+2}{3(x_{t}+1)^{3}}-\frac{2}{3} \right]
  +\frac{m_{W}}{\pi M}\bigg(-1  
   + \delta_{Db}\bigg[\frac{-x_{t}^{2}+x_{t}^{6}-4x_{t}^{4}\log{x_{t}}}{(x_{t}^{2}-1)^{3}}\bigg]\bigg) \,,
  \nl
  \hat{c}_{g}^{(2)}(\mu)&=
  \frac{\alpha_s}{4\pi}\bigg\{
 N_\ell\left( - {16\over 9} \log{\mu\over m_W} - 2 \right) - {4 (2+ 3x_t)\over 9(1+x_t)^3}\log{\mu \over m_W(1+x_t)}
\nl
&\quad
-{4 ( 12 x_t^5 - 36 x_t^4 + 36 x_t^3 - 12 x_t^2 + 3 x_t - 2)\over 9 (x_t-1)^3}\log{x_t\over 1+x_t}
- {8 x_t ( -3 + 7 x_t^2) \over 9(x_t^2-1)^3} \log 2
\nl
&\quad
- { 48 x_t^6 + 24 x_t^5 - 104 x_t^4 - 35 x_t^3 + 20 x_t^2 + 13 x_t + 18 \over 9(x_t^2-1)^2 (1+x_t)}
  \nl
  &\quad
  + \frac{m_W}{\pi M}\bigg[
  N_\ell\left( \frac83 \log{\mu\over m_W} - \frac13 \right)
 + \frac{16 x_t^4}{3(x_t^2-1)^3} \log{x_t} \log{\mu\over m_W}
 - \frac{4(3x_t^2-1)}{3(x_t^2-1)^2}     \log{\mu \over m_W}
 + \frac{16 x_t^2}{3} \log^2{x_t}
 \nl
 &\quad
 - \frac{ 4( 4x_t^6-16x_t^4+6x_t^2+1)}{3(x_t^2-1)^3} \log{x_t}
 + \frac{ 8x_t^2(x_t^6-3x_t^4+4x_t^2-1)}{3(x_t^2-1)^3} {\rm Li}_2(1-x_t^2)
 + \frac{4\pi^2 x_t^2}{9}
 \nl
 &\quad
 - \frac{8 x_t^4 - 7 x_t^2 + 1}{3(x_t^2-1)^2} 
 \bigg]
\Bigg\}
 \,. 
\end{align}
Here ${\rm Li}_{2}(z) \equiv \sum_{k = 1}^{\infty} {z^k}/{k^2}$ is the polylogarithm
of order 2. 
We also introduce the shorthand notation $c_i = (\pi\alpha_2^2/m_W^3) \hat{c}_i$ for
the effective operator coefficients, $x_i = m_i/m_W$ for masses expressed in units of $m_W$,
subscripts $U$ and $D$ denote arbitrary up-type ($u$, $c$ or $t$) or down-type ($d$, $s$ or $b$)
quarks, respectively (so that the Kronecker delta, $\delta_{Db}$,
is equal to unity for $D = b$ and vanishes for $D = d$, $s$),  
and $N_\ell=2$ is the number of massless Standard Model generations.  
The leading power results, represented by $M\to \infty$ in Eq.~(\ref{eq:results}),
were obtained in Ref.~\cite{Hill:2011be}.%
\footnote{
  In obtaining the results (\ref{eq:results}), it is important to evaluate all integrals and
  bare coefficients in $d=4-2\epsilon$ dimensions~\cite{Hill:2011be,Hill:2014yka}.  For a related
  discussion see Ref.~\cite{Weinzierl:2014iaa}. 
}
Let us remark that our results (\ref{eq:results}) obey the correct formal limit at
small $x_t$:~\cite{Hill:2011be}
\begin{align}
  c^{(0)}_g |_{x_t\to 0} &= c^{(0)}_g ( n_f=6)  - {\alpha_s \over 12\pi} c^{(0)}_q(n_f=6) + \order(\alpha_s^2) \,,
  \nl
  c^{(2)}_g |_{x_t\to 0} &= c^{(2)}_g ( n_f=6) - {\alpha_s \over 3\pi} \log{m_t\over \mu} c^{(2)}_q(n_f=6)
  + \order(\alpha_s^2) \,,
\end{align}
where $c(n_f=6)$ denotes the coefficient in six-flavor QCD computed with three massless generations
(i.e., $m_t \ll m_W$).%
\footnote{
  In particular, the quark matching coefficients are
  $\hat{c}_q^{(0)}(n_f=6)=-{1\over x_h^2}- {m_W\over \pi M} {c_H \over \alpha_2^2 x_h^2}$ and 
  $\hat{c}_q^{(2)}(n_f=6)=\frac23 - {m_W\over \pi M}$ for $q=u,d,c,s,t,b$. 
  The gluon matching coefficients are obtained by omitting the top quark loop contributions
  in Eq.~(\ref{eq:results}) and setting $N_\ell=3$: 
  $\hat{c}_g^{(0)}(n_f=6)= {\alpha_s\over 8\pi}$ and 
  $\hat{c}_g^{(2)}(n_f=6) = {\alpha_s\over 4\pi}\left[ -{16\over 3}\log{\mu\over m_W} - 6
    + {m_W\over \pi M}\left( 8 \log{\mu\over m_W} - 1 \right) 
    \right]$. 
}
At large $x_t$, $m_t \gg m_W$,
the top quark contributions to the coefficients are of order $\sim m_W^2/m_t^2$. 
For the special case of a Majorana
fermion ($c_H = -3\alpha_2^2$), the $1/M$ corrections for $c^{(0)}_{q,g}$ and $c^{(2)}_q$
are reproduced by an expansion of expressions in Ref.~\cite{Hisano:2015rsa}.  However,
already at leading power
the expression in Ref.~\cite{Hisano:2015rsa} for $c^{(2)}_g$ 
disagrees with the corresponding results in Ref.~\cite{Hill:2011be} and
Eq.~(\ref{eq:results}).
We note that the expression for $c^{(2)}_g$ in Ref.~\cite{Hisano:2015rsa} does
not have the correct $m_t\to 0$ limit.

\section{Cross sections \label{sec:numerics}}

Let us consider the standard benchmark process for direct detection:
the zero velocity limit of (spin-independent) WIMP-nucleon
scattering. The cross section is determined by the spin-0 and spin-2
matrix elements, $\mathcal{M}^{(0)}_{N}$ and $\mathcal{M}^{(2)}_{N}$,
of the operators in Eq.~(\ref{eq:op}),
\begin{equation}
\mathcal{M}^{(S)}_{N} = \sum_{i= q,g} c^{(S)}_{i}(\mu_{0}) \langle N | O^{(S)}_{i}(\mu_{0})|N\rangle \,.
\end{equation}
In order to evaluate the hadronic matrix elements using available
low energy inputs, the five flavor
QCD theory must be matched to the appropriate three or four flavor
theory, accounting for heavy quark threshold matching corrections and 
renormalization group evolution from electroweak to hadronic scales. 
Details of this matching can be found in Ref.~\cite{Hill:2014yxa}.
For the spin-0 matrix elements, we match to the three flavor theory with NNNLO
QCD corrections,%
\footnote{
  For the leading power analysis, this corresponds to amplitude ``5''
  discussed in Figure~2 and Section~6.2.3 of Ref.~\cite{Hill:2014yxa}.
  }
and following Ref.~\cite{Hill:2014yxa} make the default scale choices
$\mu_{t} = (m_t+m_W)/2 = 126\,{\rm GeV}$, $\mu_{b} = 4.75\,{\rm GeV}$, 
$\mu_{c} = 1.4\,{\rm GeV}$, and $\mu_{0} = 1.2\,{\rm GeV}$. 
For the spin-2 matrix elements, we use NLO running and matching, and
check that our evaluation is consistent with an evaluation at the
weak scale, in the five flavor theory.
The impact of higher order perturbative QCD corrections
is estimated by varying factorization scales
$m_W^2/2 \le \mu_t^2 \le 2 m_t^2$,
$m_b^2/2 \le \mu_b^2 \le 2 m_b^2$,
$m_c^2/2 \le \mu_c^2 \le 2 m_c^2$,
and
$1.0\,{\rm GeV} \le \mu_0 \le 1.4\,{\rm GeV}$.
There are additional uncertainties
associated with the hadronic form factors that characterize the
overlap between the nucleon states and the quark and gluon
operators.
We employ the form factor central values and uncertainties
from Ref.~\cite{Hill:2014yxa}, which were adapted from
Refs.~\cite{Gasser:1982ap,Durr:2011mp,Junnarkar:2013ac,Martin:2009iq}
(see also Ref.~\cite{Crivellin:2013ipa}). 
Errors from all sources are added in quadrature to obtain the
total cross section error.

\begin{figure}[t]
\begin{center}
\includegraphics[width=8.5cm]{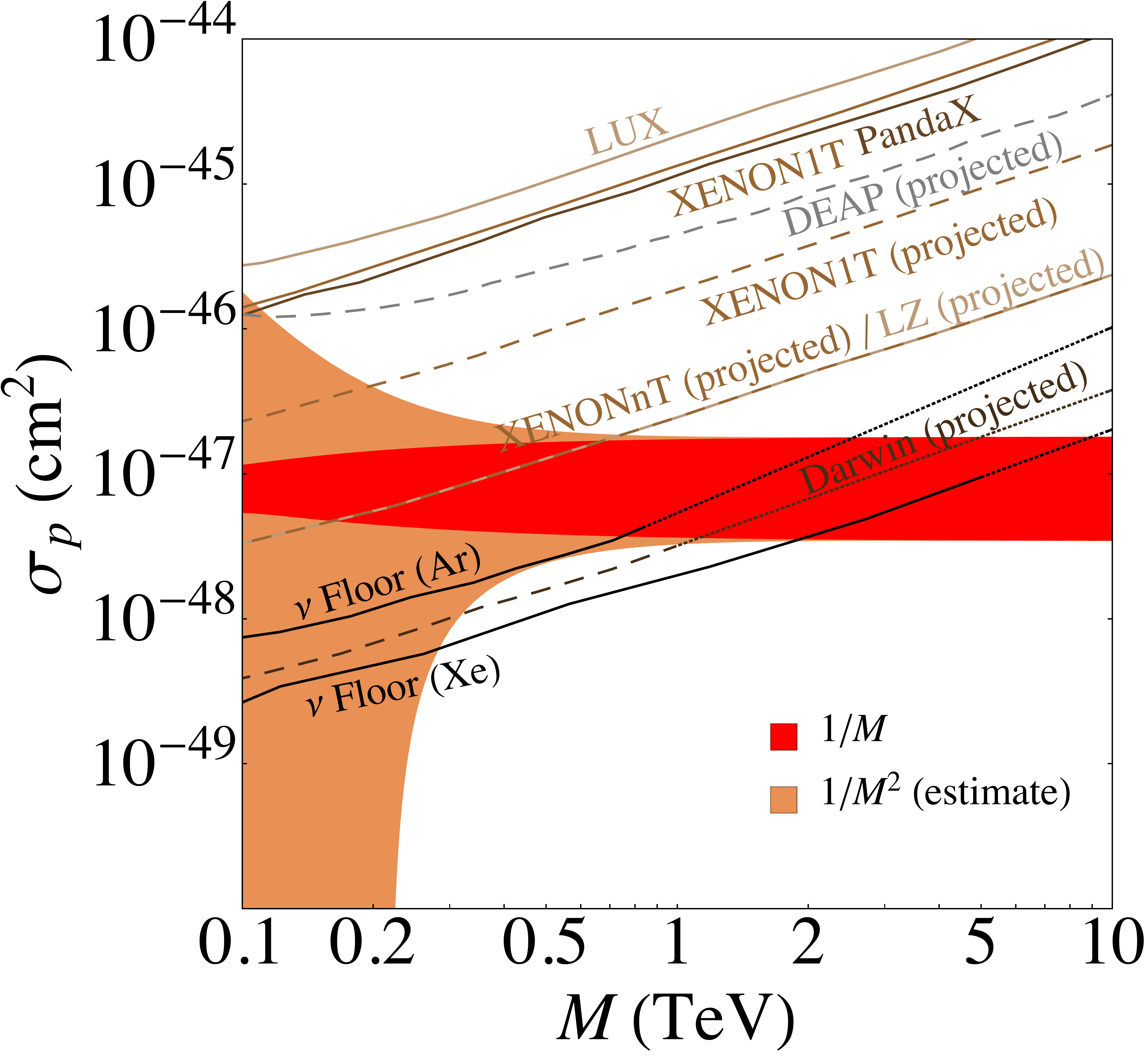}\hspace{0.2cm}
\includegraphics[width=8.5cm]{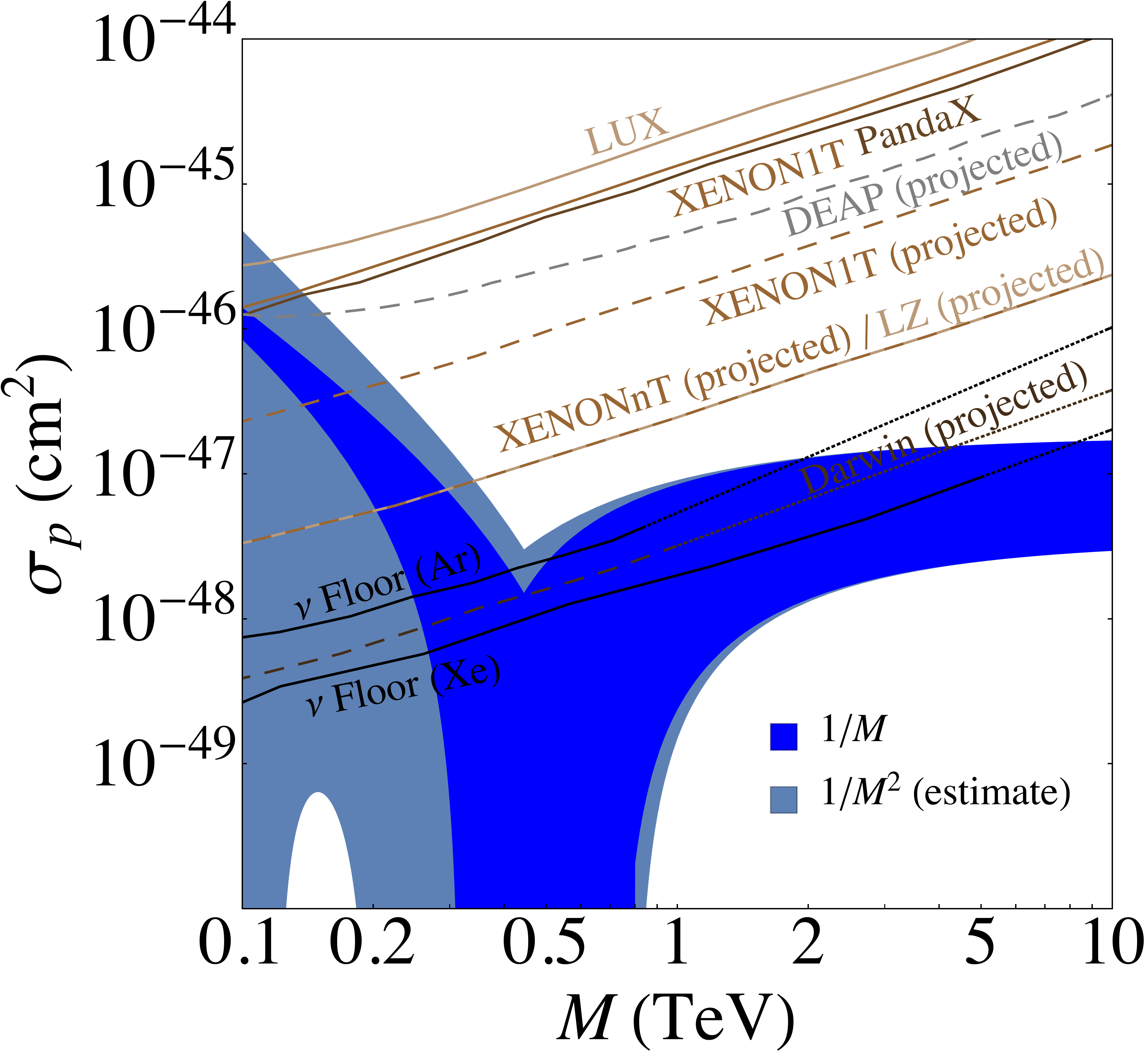}
\end{center}
\caption{The WIMP-proton scattering cross section as a function of
  WIMP mass $M$ for a Majorana WIMP (left panel) and a scalar WIMP
  (right panel), which correspond to the $c_H$ values in
  Eqs.~(\ref{eq:cH_T}) and (\ref{eq:cH_S}), respectively. The inner band
  is the cross section obtained from the scalar and
  tensor amplitudes computed through $\order(1/M)$. The outer band
  includes an estimate for the $\order(1/M^{2})$
  contributions. The neutrino floor for both Argon and Xenon direct
  detection experiments are from Ref.~\cite{Billard:2013qya}, and are
  shown by black solid lines; our extrapolation to larger masses is
  denoted with black dashed lines. Also shown with solid lines are the
  current bounds from LUX~\cite{Akerib:2016vxi},
  XENON1T~\cite{Aprile:2017iyp}, and PandaX-II~\cite{Cui:2017nnn}.
  Projected sensitivities of future experiments are shown with dotted
  lines: DEAP-3600~\cite{Amaudruz:2017ekt},
  XENON1T and XENONnT~\cite{Aprile:2015uzo}, LZ~\cite{Akerib:2015cja},
   and DARWIN~\cite{Aalbers:2016jon}.
}
\label{fig:sigmaVsM}
\end{figure}

Neglecting numerically small CKM factors and isospin violation in
nucleon matrix elements~\cite{Hill:2014yxa}, the cross sections
for scattering on protons or neutrons are identical:%
\footnote{
  The Wilson coefficients $c^{(S)}_{u}$ and $c^{(S)}_{d}$ in Eq.~(\ref{eq:results})
  are identical.  The light quark operators in Eq.~(\ref{eq:weaklagrangian})
  thus appear in the combinations $O^{(S)}_u + O^{(S)}_d$, whose
  proton and neutron matrix elements are identical up to isospin violating
  corrections.  These percent level corrections, 
  proportional to $\alpha \approx 1/137$ or $(m_u-m_d)/\Lambda_{\rm QCD}$,
  are subdominant in the error budget for ${\cal M}_N^{(S)}$.  See Ref.~\cite{Hill:2014yxa}
  for details. 
}
\begin{equation}
\sigma_p \approx \sigma_n = \frac{m_{r}^{2}}{\pi}|\mathcal{M}^{(0)}_{p} + \mathcal{M}^{(2)}_{p}|^{2} \,,  
\end{equation}
where $m_r = m_p M /(m_p + M) \approx m_p$ is the reduced mass of
the WIMP-nucleon system. 
In Fig.~\ref{fig:sigmaVsM} we show the cross section including
first order power corrections as a function of $M$
for a fundamental fermion, Eq.~(\ref{eq:cH_T}), and for a composite scalar,
Eq.~(\ref{eq:cH_S}).
The central value amplitudes, in units with ${\cal M}^{(2)}_p|_{M\to\infty} = 1$,
are
\begin{align}\label{eq:amp_num}
  {\cal M}_p^{(2)} = 1 - 0.52 {m_W \over M} \,, \qquad
  {\cal M}_p^{(0)} =  -0.81 - 0.50 {c_H \over 3\alpha_2^2}  {m_W\over M} \,. 
\end{align}
The numerical evaluation (\ref{eq:amp_num}) exhibits the partial cancellation
of the universal $M\to \infty$ result. For the Majorana fermion case,
where $c_H=-3\alpha_2^2$, the $m_W/M$ power correction also exhibits a
surprising cancellation. 
The impact of neglected higher-order power corrections is estimated
by including an uncertainty in the tensor amplitude as
${\cal M}_p^{(2)} \propto {\cal M}_p^{(2)}|_{M\to \infty} \left[ 1 \pm (m_W/M)^2 \right]$.
At large mass, the power corrections vanish, and the universal result
with central value and uncertainty from Ref.~\cite{Hill:2014yxa} is
reproduced.
At finite WIMP mass, the dependence of the cross section on the
Higgs coupling $c_H$ differentiates the fermion and scalar cases. 

Figure~\ref{fig:sigmaVsM} compares to existing limits from LUX
\cite{Akerib:2016vxi}, XENON1T \cite{Aprile:2017iyp}, and PandaX-II
\cite{Cui:2017nnn},%
\footnote{
  For masses larger than the ranges reported in these references,
  we have displayed an extrapolation assuming simple scaling
  with the WIMP number abundance, $\sigma_{\rm limit} \propto M$. 
  }
and to projected sensitivities for the
Xenon based experiments XENONnT~\cite{Aprile:2015uzo}, LZ~\cite{Akerib:2015cja}, and
DARWIN~\cite{Aalbers:2016jon}, and the
Argon based experiment DEAP-3600 \cite{Amaudruz:2017ekt}.
Also shown is the ``discovery limit'' for both Xenon and Argon
due to neutrino backgrounds, taken from Ref.~\cite{Billard:2013qya}.

\section{Summary \label{sec:summ}}

The scattering of atomic nuclei from approximately static sources of
electroweak SU(2) is a well posed but intricate field theory problem
that finds application in the search for WIMP dark matter in our
local halo.
LHC bounds have pushed the scale of new physics into a regime of
large mass where direct detection is more challenging; however at the same time,
universal predictions emerge in this regime and provide well-defined targets for
next generation searches.

Generic amplitude level cancellations imply a potentially enhanced sensitivity
of direct detection rate predictions to naively power suppressed interactions.
In this paper we considered the general framework to analyze these power
corrections, and analyzed the canonical case of a self-conjugate
electroweak-triplet WIMP through order $1/M$.
Owing to heavy particle universality, the leading cross section prediction
is identical whether such a WIMP is fermion or scalar, elementary or composite,
and whether the WIMP is accompanied by other, heavier, particles in the
Standard Model extension.
Power corrections differentiate these scenarios, as
illustrated in Fig.~\ref{fig:sigmaVsM} for the benchmark low-velocity WIMP-nucleon cross section.
For the elementary fermion case, two contributions to the power correction
largely cancel, resulting in a small deviation from the universal $M\to\infty$
limit.  Our result represents the most complete calculation of
the cross section for wino-like dark matter in the TeV regime.
A standard thermal cosmology, consistent with the observed dark matter abundance,
predicts $M \sim 2-3\,{\rm TeV}$ for such electroweak charged
WIMPs~\cite{Hisano:2004ds,Hisano:2006nn,Cirelli:2007xd,Hryczuk:2010zi,Hryczuk:2011vi}. 
The elementary Majorana fermion case involves no free parameters, and a 
prediction $M \approx 2.9\,{\rm TeV}$ is obtained after careful
accounting for nonperturbative enhancements~\cite{Beneke:2016ync}. 
For the scalar case, the precise annihilation cross section, and hence cosmological
mass constraint, depends on internal structure.
At the TeV mass scales indicated by cosmological arguments, the
predicted WIMP-nucleus scattering rate is comparable to the rate for
neutrino-induced backgrounds. 
This cross section benchmark motivates very large scale detectors,
and techniques to understand and probe into the so-called
neutrino floor~\cite{Mayet:2016zxu}.

A number of investigations are suggested by our results.  Besides its
computational power, the heavy WIMP
expansion provides an excellent classification scheme for WIMP direct
detection in the increasingly important heavy WIMP regime. 
The SU(2) triplet (i.e., wino-like) case 
represents a canonical benchmark.  Other quantum numbers such as the
higgsino-like case may be similarly investigated.
The proximity of the triplet cross section in Fig.~\ref{fig:sigmaVsM} to
the neutrino floor makes the
precise WIMP mass of particular interest.  For the composite scalar case,
new nonperturbative physics enters in two key places:
the Higgs coupling parameter $c_H$ that determines the size of the direct
detection cross section; and the annihilation process that determines
the cosmological mass constraint within a specified cosmological model.
This physics could be accessed by
lattice field theory~\cite{Appelquist:2014jch} and/or chiral lagrangian
analysis for the new strongly coupled sector. 
Nuclear effects such as two-body correlations could potentially have differing impacts on the 
spin-0 and spin-2 operators in Eq.~(\ref{eq:weaklagrangian}).  Like
the $1/M$ corrections, the existence of a severe cancellation in the
leading cross section can potentially enhance the impact of such naively
subleading effects. Existing estimates for such nuclear effects, focused
on the spin-0 sector, indicate
a small impact relative to other
uncertainties~\cite{Cirigliano:2012pq,Hoferichter:2016nvd,Korber:2017ery},
however a more systematic analysis is warranted. 

\vskip 0.2in
\noindent
{\bf Acknowledgements}
\vskip 0.1in
\noindent
CYC would like to thank C. Burgess for helpful discussions.
RJH thanks TRIUMF and Perimeter Institute for hospitality. 
Work of MPS was supported by the Office of High Energy Physics of the U.S. DOE under
Contract Numbers DE-AC02-05CH11231  and  DE-SC0011632.
Research at the Perimeter Institute is supported in part by the Government of Canada through NSERC
and by the Province of Ontario through MEDT.
TRIUMF receives federal funding via a contribution agreement with the National Research Council of Canada.
Fermilab is operated by Fermi Research Alliance, LLC under Contract No. DE-AC02-07CH11359 with the United States
Department of Energy. 

%\end{fmffile} 

\bibliography{HWET_1overM}

\end{document}